\documentstyle[12pt,aps,preprint,epsf]{revtex}
\def\whatjournal{N}
\newlinechar=`\^^J

\def\ordernpb#1#2#3{{\bf#1} (#3) #2}
\if P\whatjournal {\global\def\order#1#2#3{\orderprd{#1}{#2}{#3}}}
                    \immediate\write16{^^J PRD references ^^J}\else
                   {\global\def\order#1#2#3{\ordernpb{#1}{#2}{#3}}}
                    \immediate\write16{^^J NPB references ^^J}
\fi
\def\ijmpa#1#2#3{{\rm Int. J. of Mod. Phys. {\bf A}}\order{#1}{#2}{#3}}
\def\plb#1#2#3{{\rm Phys. Lett. {\bf B}}\order{#1}{#2}{#3}}
\def\npb#1#2#3{{\rm Nucl. Phys. {\bf B}}\order{#1}{#2}{#3}}
\def\physa#1#2#3{{\rm Physica {\bf A}}\order{#1}{#2}{#3}}
\def\zphys#1#2#3{{\rm Z. Phys. {\bf C}}\order{#1}{#2}{#3}}

\def\sm{Standard Model}
\def\acal{{\cal A}}
\def\lcal{{\cal L}}
\def\inv#1{{1\over#1}}
\def\ocal{{\cal O}}
\def\su#1{{SU(#1)}}
\def\ui{U(1)}
\def\vev{vacuum expectation value}
\def\gev{\hbox{GeV}}
\def\half{{1\over2}}
\def\up#1{^{\left( #1 \right) }}
\def\lesim{\,{\raise-3pt\hbox{$\sim$}}\!\!\!\!\!{\raise2pt\hbox{$<$}}\,}
\def\gesim{\,{\raise-3pt\hbox{$\sim$}}\!\!\!\!\!{\raise2pt\hbox{$>$}}\,}
\def\tev{\hbox{TeV}}
\def\qwe{b}
\def\vacpol{vac. pol.}
\def\chir{{\rm chir}}
\def\lin{{\rm lin}}
\def\phichir{\phi_\chir}
\def\Delchir{\Delta_\chir}
\def\Dellin{\Delta_\lin}
\def\svacpol{S_{\rm\vacpol}}
\def\tvacpol{T_{\rm\vacpol}}
\def\uvacpol{U_{\rm\vacpol}}
\def\qweaa{ x \qwe_{\phi W}+(1-x)\qwe_{\phi B}-2y \qwe_{W B} }
\def\qwezz{ x \qwe_{\phi B}+(1-x)\qwe_{\phi W}+2y \qwe_{W B} }

\begin{document}

\title{Effective operator contributions to the oblique parameters}

\author{
G.~S\'anchez-Col\'on\footnote{
\baselineskip 12 pt
Permanent address:
Departamento de F\'{\i}sica Aplicada.
Centro de Investigaci\'on y de Estudios Avanzados del IPN.
Unidad M\'erida.
A.P. 73, Cordemex.
M\'erida, Yucat\'an 97310, MEXICO.
}\ and J.~Wudka\
}
\address{
Physics Department, University of California Riverside\\
Riverside, CA 92521-0413, U.S.A.\\
}

\preprint{UCRHEP-T229}

\date{\today}

\maketitle

\begin{abstract}
We present a model and process independent study of the contributions
from non-\sm\ physics to the oblique parameters $S$, $T$ and $U$.
We show that within an
effective lagrangian parameterization the expressions for the oblique
parameters in terms of observables are consistent, while those in terms
of the vector-boson vacuum polarization tensors are ambiguous. We obtain
the constraints on the scale of new physics derived from current data on
$S$, $T$ and $U$ and note that deviations in $U$ from its \sm\ value
would favor a scenario where the underlying physics does not decouple.
\end{abstract}

\newpage

The oblique parameters $S$, $T$ and $U$~\cite{peskin} are known to be
sensitive probes of non-\sm\ physics. Because of this they are often
used in deriving bounds on the scale (and other properties) of new
physics form the existing and expected experimental bounds~\cite{new.stu}.
These
parameters are often defined in terms of the vector-boson vacuum
polarization tensors~\cite{peskin}, but a practical definition requires
them to be expressed in terms of direct observables~\cite{roy}. Thus,
whenever the contributions from new interactions to the oblique
parameters are calculated, the modifications to all observable
quantities involved should be included.

When dealing with specific models the calculation of the oblique
parameters is a straight-forward exercise. In contrast, when
considering the same calculation using a model-independent (effective
Lagrangian) parameterization of the new-physics effects, some subtleties
arise. The reason is that the effective Lagrangian parameterization is
not unique in the sense that one can change the effective Lagrangian
without affecting the S matrix~\cite{arzt}, yet these modifications do
alter the vector boson vacuum polarization and, with this, the
corresponding definition of the oblique parameters. Because of
this the definition of $S$, $T$ and $U$ in terms of the the vector-boson
vacuum polarization tensors is ambiguous.

We will show that this problem can be avoided by defining $S$, $T$ and
$U$ in terms of observables (which unfortunately is seldom the
case~\cite{hagiwara,wudka}). We will obtain the complete expressions for
the contributions from non-\sm\ physics to the oblique parameters within
an effective Lagrangian parameterization. From these expressions
unambiguous limits on the scale of new physics can be derived.  Finally
we will also argue that accurate measurements of the $U$ parameter will
provide information on whether the heavy physics decouples.

Within the \sm\ the oblique parameters vanish at tree-level, but they are
non-zero
at one loop~\cite{sm.stu}; new physics will also, in general, generate
non-vanishing
contributions~\cite{new.stu}.
To lowest order we expect $S=\delta_{\rm rad} S+\Delta S$ (with similar
expressions
for $T$ and $U$), where $\delta_{\rm rad} S$ denotes the radiative \sm\
contributions
and $\Delta S$ the contributions generated by the heavy physics.
The quantities $\delta_{\rm rad} (S,T,U)$ are well known and have been studied
extensively~\cite{sm.stu}.
In this paper we concentrate on $\Delta (S,T,U)$ keeping in mind that these
quantities denote the deviations from the \sm\ predictions with radiative
corrections
included; in calculating $ \Delta S, \; \Delta T, \; \Delta U$ we will ignore
all
\sm\ loop effects.

The oblique parameters can be expressed~\cite{roy} in terms of
the fine-structure constant $\alpha$ (measured at the $Z$ mass),
the vector boson masses $M_Z$ and $M_W$, the Fermi constant $G_F$,
and the width $\Gamma(Z\to\bar{l}^+l^-)$ and forward-backward
asymmetry, $\acal_{FB}(Z\to\bar{l}^+l^-)$, for the decay of the $Z$ into
charged leptons.
{}From $ \acal_{FB} $ we obtain $g_V/g_A$ (the vector-coupling
to axial-coupling ratio of the $Z$ to the charged leptons); with
this result and using the other observables the oblique parameters are obtained
from
\begin{eqnarray}
\Gamma ( Z \rightarrow \bar \ell \ell ) &=& { G_F M_Z^3 \over 24 \sqrt{2}
\; \pi }
\left(1 + { g_V^2
\over g_A^2 } \right) \left( 1 + \alpha T \right)
 ,\nonumber\\
{1 \over 4} \left(1 + { g_V \over g_A } \right) &=& s_0^2 - { s_0^2 (1-s_0^2)
\over 1-2 s_0^2 } \alpha T + { \alpha S \over 4(1-2 s_0^2) } , \nonumber\\
{ M_W^2 \over M_Z^2 ( 1 - s_0^2) } &=& 1 + { 1-s_0^2 \over 1-2s_0^2 } \alpha T
-{ \alpha S \over 2 (1-2 s_0^2) } + {\alpha U \over 4 s_0^2 },
\label{stu.physical}
\end{eqnarray}
where
\begin{equation}
s^2_0 ( 1 - s^2_0 ) = { \pi \alpha \over \sqrt{2}G_F M^2_Z } .
\end{equation}

To obtain the heavy physics contributions to the oblique parameters it is then
necessary to determine the contributions to the observables used in the above
definitions.
In this paper we will use an effective Lagrangian parameterization of the heavy
physics~\cite{wudka,leff.refs} which has the advantage of being model and
process independent.
The detailed form of the effective Lagrangian depends crucially on the low
energy spectrum, we will include three families of fermions as well as the
usual \sm\ gauge bosons. For the scalars we will consider two possibilities: in
the first, which we label the {\em linear case}~\cite{leff.linear}, we assume a
single light scalar doublet; in the second, which we call the {\em chiral
case}~\cite{leff.chiral}, we assume that there are no light physical scalars.
In both cases we denote the scale of new physics by $\Lambda$.

For the linear case the part of the effective Lagrangian which
contributes to the oblique parameters takes the
form~\cite{leff.linear,bw}
\begin{equation}
\lcal_{\rm eff} = \inv{\Lambda^2 } \sum_i \qwe_i \ocal_i + O \left( \inv{
\Lambda^3 } \right)
\label{leff}
\end{equation}
where~\footnote{
We use the following  conventions~\cite{bw}: $I,J,K$ denote $\su2$ indices, the
Pauli matrices are denoted by $\tau^I$; $W_\mu^I$ and $B_\mu$ denote the $\su2$
and $\ui$ gauge fields, and $W_{\mu\nu}^I$ and  $B_{\mu\nu}$ the corresponding
curvatures; the gauge coupling constants are denoted by $g$ and $g'$
respectively.
Left-handed quark and lepton doublets are denoted by $q$ and $\ell$
respectively; right-handed up and down-type quarks correspond to $u$ and $d$,
while the right-handed charged lepton corresponds to $e$; all fermion fields
have implicit family indices. The scalar doublet is denoted by $\phi$ and the
covariant derivative by $ D_\mu $. The scalar \vev\ is denoted by $v$ defined
so that $ v\simeq 246 \gev $.}
\begin{equation}
\begin{array}{lll}
\ocal_{ \phi W } = \half \left( \phi^\dagger \phi \right) W_{\mu \nu }^I W^{I
\, \mu \nu }
& \quad
\ocal_{ \phi B } = \half \left( \phi^\dagger \phi \right) B_{\mu \nu } B^{ \mu
\nu }
& \quad
\ocal_{ W B } = \left( \phi^\dagger \tau^I \phi \right) W_{\mu \nu }^I B^{ \mu
\nu }
\\
\ocal_\phi\up1 = \left( \phi^\dagger \phi \right) \left[ \left( D_\mu \phi
\right)^\dagger D^\mu \phi \right]
& \quad
\ocal_\phi\up3 = \left( \phi^\dagger D^\mu \phi \right) \left[ \left( D_\mu
\phi \right)^\dagger \phi \right]
& \quad
\ocal_{\phi \ell}\up1 = i \left( \phi^\dagger D^\mu \phi \right) \left( \bar
\ell \gamma_\mu \ell \right)
\\
\ocal_{\phi \ell}\up3 = i \left( \phi^\dagger \tau^I D^\mu \phi \right) \left(
\bar \ell \tau^I \gamma_\mu \ell \right)
& \quad
\ocal_{\phi e} = i \left( \phi^\dagger D^\mu \phi \right) \left( \bar e
\gamma_\mu e \right)
& \quad
\ocal_{\ell \ell }\up3 = \half \left( \bar \ell \tau^I \gamma^\mu \ell \right)
\left( \bar \ell \tau^I \gamma_\mu \ell \right).
\end{array}
\label{ops}
\end{equation}
The coefficients $ \qwe_i $ parameterize all heavy physics contributions to the
oblique parameters. The choice of operators is, however, not
unique~\cite{arzt}; we will discuss this issue below.

When there are no-light scalars (chiral case) the effective Lagrangian
can be obtained from (\ref{leff}) by replacing
\begin{equation}
\phi\rightarrow\phichir =\Sigma\left(
\begin{array}{c}
0 \\
v \\
\end{array}
\right) ; \qquad \Sigma^\dagger \cdot \Sigma = 1 
\label{chir}
\end{equation}
where $v\simeq 246\gev$.
In this case the effects of the operators $\ocal_{\phi W}$,
$\ocal_{\phi B}$ and $\ocal_\phi\up1$ can be absorbed by an
appropriate renormalization of the \sm\ parameters.
Therefore, in the chiral case we set
$\qwe_{\phi W}=\qwe_{\phi B}=\qwe_\phi\up1=0$.
A consistent expansion of the chiral effective Lagrangian~\cite{wudka}
requires that we include all operators if $f$ fermion fields and $d$
derivatives are such that~\footnote{This is a generalization of the
derivative expansion when fermions are present.}
$d+f/2\le 4$, hence we must also include the operator~\cite{appelquist.wu}
\begin{equation}
\ocal_{WW} = \half \left( \phichir^\dagger \tau^I \phichir \right)
\left( \phichir^\dagger \tau^J \phichir \right) W_{\mu \nu}^I W^{J \,
\mu \nu}
\label{extra}
\end{equation}
(which does not appear in
(\ref{ops}) since in the linear case it corresponds to a dimension 8
operator which will generate subdominant contributions to the
oblique parameters).

Using (\ref{leff}, \ref{ops}) (together with (\ref{chir},
\ref{extra}) in the chiral case) we obtain the heavy physics contributions to
the
$Z$ couplings and mass, the $W$ mass, $ \alpha $ and $ G_F $. We
first provide the expressions in terms of the $\su2$ and
$\ui $ gauge coupling constants, $ g $ and $g' $ respectively, and the \vev\
$v$; and then express these in terms of direct observables.

We consider first the case where there is a single light scalar doublet
(the linear case).
The $Z$ axial and vector couplings to the charged leptons equal,
respectively
\begin{equation}
g_A = g_R-g_L, \qquad g_V = g_R+g_L,
\label{def.ga.gv}
\end{equation}
where
\begin{eqnarray}
g_L &=& - \half + x - \half \left[
 (1-x)(1 + 2x) \qwe_{ \phi W}
- (x + 2 y^2)  \qwe_{ \phi B }
- 2 y (1 - 2x) \qwe_{ W B }
+              \qwe_{ \phi \ell }\up1
+              \qwe_{ \phi \ell }\up3
                            \right] \epsilon + O(\epsilon^2); \nonumber \\
g_R &=&  x - \half \left[
              \qwe_{ \phi e}
+ 2 y^2       \qwe_{ \phi W }
- 2 x (2 - x) \qwe_{ \phi B }
- 4 y (1 - x) \qwe_{ W B }
                   \right]  \epsilon + O(\epsilon^2);
\end{eqnarray}
and
\begin{equation}
\epsilon = \half { v^2 \over \Lambda^2 }, \qquad
x = { g'{}^2 \over g^2 + g'{}^2 }, \qquad
y = { g' \; g \over g^2 + g'{}^2 } .
\label{def.x.y.eps}
\end{equation}

For the remaining observables we find
\begin{eqnarray}
M_W &=& { g \; v \over 2 } \left[ 1 + \left( \qwe_{\phi W} + \half
\qwe_\phi\up1 \right) \epsilon \right] + O(\epsilon^2); \nonumber \\
M_Z &=& { g \; v \over 2 \; \sqrt{1-x}} \left\{ 1 + \left[ \qwezz +
\qwe_\phi\up1 + \half \qwe_\phi\up3 \right] \epsilon \right\} + O(\epsilon^2);
\nonumber \\
G_F &=& \inv{ \sqrt{2} \; v^2} \left[1 + \left( 2 \qwe_{\ell \ell} \up3 + 4
\qwe_{\phi \ell}\up3 - \qwe_\phi\up1 - 4 \qwe_{\phi W } \right) \epsilon
\right] + O(\epsilon^2); \nonumber \\
\alpha &=& { g^2 x \over 4 \pi } \left\{ 1 + 2 \left[ \qweaa \right] \epsilon
\right\} + O(\epsilon^2);
\label{def.mw.mz.gf.alp}
\end{eqnarray}
which can be used to express $x$, $g$, $\epsilon $, etc. in terms of
observables.
These expressions do {\em not} contain the \sm\ radiative corrections
since, as discussed above, we expect the corresponding contributions to
the oblique parameters to be additive and we are interested only in the
contributions generated by the heavy physics. Substituting
(\ref{def.ga.gv}-\ref{def.mw.mz.gf.alp}) into (\ref{stu.physical}) we get
\begin{eqnarray}
\Dellin T &=& - \frac{4\pi}{g^2 \; x} \left( 2\qwe_{\ell \ell}\up3 + 2
\qwe_{\phi \ell}\up3 - 2\qwe_{\phi \ell}\up1+ \qwe_\phi\up3 + 2\qwe_{\phi e} -4
\qwe_{\phi W} \right) \epsilon + O ( \epsilon^2 ), \nonumber\\
\Dellin S &=& \frac{8\pi}{g^2 \; x} \left[ -\qwe_{\phi e} + 2 x\left(\qwe_{\phi
\ell}\up1 + \qwe_{\phi \ell}\up3 \right) + 4 y \qwe_{ W B } \right] \epsilon +
O ( \epsilon^2), \nonumber \\
\Dellin U &=& \frac{16\pi}{g^2} \left( 2 \qwe_{\ell \ell}\up3 + 2\qwe_{\phi
\ell}\up3 - 2 \qwe_{\phi \ell}\up1 + \qwe_{\phi e} - 4 \qwe_{ \phi W } \right)
\epsilon + O ( \epsilon^2).
\label{stu}
\end{eqnarray}

In the chiral case, using (\ref{chir}) and (\ref{extra}) we obtain
\begin{eqnarray}
\Delchir T &=& - \frac{4\pi}{g^2 \; x} \left( 2\qwe_{\ell \ell}\up3 + 2
\qwe_{\phi \ell}\up3 - 2\qwe_{\phi \ell}\up1+ \qwe_\phi\up3 + 2\qwe_{\phi e}
\right) \epsilon + O ( \epsilon^2 ), \nonumber\\
\Delchir S &=& \frac{8\pi}{g^2 \; x} \left[ -\qwe_{\phi e} + 2
x\left(\qwe_{\phi \ell}\up1 + \qwe_{\phi \ell}\up3 \right) + 4 y \qwe_{ W B }
\right] \epsilon + O ( \epsilon^2), \nonumber \\
\Delchir U &=& \frac{16\pi}{g^2} \left( 2 \qwe_{\ell \ell}\up3 + 2\qwe_{\phi
\ell}\up3 - 2 \qwe_{\phi \ell}\up1 + \qwe_{\phi e} - 2 \qwe_{ W W } \right)
\epsilon + O ( \epsilon^2).
\label{stu.chiral}
\end{eqnarray}
where $ \qwe_{WW} $ is the coefficient of $ \ocal_{WW} $ in (\ref{extra})
(which has dimension 4).

The above expressions can be re-written in terms of observables using
the tree-level relations
\begin{equation}
g^2 = 4 \sqrt{2} \, G_F M_W^2, \qquad x = 1 - (M_W/M_Z)^2 , \qquad
\epsilon = {1 \over \sqrt{8} \, G_F \Lambda^2 } .
\end{equation}

These expressions were obtained using the effective Lagrangian (\ref{leff},
\ref{ops}), together with (\ref{chir}, \ref{extra}) in the chiral case.
But it is well known~\cite{arzt} that
there is no unique choice of operators in an effective Lagrangian
parameterization.
Given two
operators $ \ocal_1 $ and $ \ocal_2 $ such that $ \ocal_1 - \ocal_2
$ vanishes when the classical equations of motion are imposed, then the
term $ \qwe_1 \ocal_1 + \qwe_2
\ocal_2 $ in the effective Lagrangian generates modifications to the
S matrix which depend {\em only} on $ \qwe_1 + \qwe_2
$~\cite{arzt}, but not on $ \qwe_1 $ and $ \qwe_2 $
independently.

Our expressions for $ \Delta S $, $ \Delta T $ and
$ \Delta U $ satisfy this property.
As an example consider the operator
\begin{equation}
\ocal_{DW}=\left(D_{\mu}W_{\nu\rho}\right)^I\;\left(D^{\mu}W^{\nu\rho}
\right)^I,
\label{odw}
\end{equation}
which, up to terms which vanish when the
classical equations of motion are imposed, satisfies
\begin{equation}
\ocal_{DW} = 2 g \ocal_W + { g^2 \over 2 } \left[ 6
\ocal_\phi\up1 + 2 m^2 \left( \phi^\dagger \phi \right)^2 - 6 \lambda
\ocal_\phi + 4 \ocal_{ \phi \ell}\up3 + 4 \ocal_{\phi q }\up3 + 2
\ocal_{\ell \ell}\up3 + 2 \ocal_{\ell q}\up3 + 2 \ocal_{q q}\up3
\right]
\end{equation}
where $m$ denotes the scalar mass, $ \lambda $ the scalar self-coupling
and where the operators not defined in (\ref{ops}) are
\begin{equation}
\begin{array}{ll}
\ocal_\phi = \inv3 \left( \phi^\dagger \phi \right)^3
& \qquad
\ocal_{\phi q}\up3 = i \left( \phi^\dagger \tau^I D^\mu \phi \right) ( \bar q
\tau^I \gamma_\mu q )
\\
\ocal_{\ell q} \up3 = ( \bar \ell \tau^I \gamma^\mu \ell ) ( \bar q \tau^I
\gamma_\mu q )
& \qquad
\ocal_{qq}\up3 = \half ( \bar q \tau^I \gamma^\mu q ) ( \bar q \tau^I
\gamma_\mu q)
\\
\ocal_W = \varepsilon_{I J K } W^I{}_\mu^\nu W^J{}_\nu^\lambda
W^K{}_\lambda^\mu.
&
\end{array}
\end{equation}
It then follows that the replacement
\begin{equation} \lcal_{\rm eff} \rightarrow
\lcal_{\rm eff} + { \qwe_{DW} \over \Lambda^2 }\ocal_{DW}
\label{repl}
\end{equation}
is equivalent to
$ \qwe_i \rightarrow \qwe_i + \delta \qwe_i $ where
\begin{eqnarray}
 \inv3 \delta \qwe_\phi\up1 =
-\inv3 \delta \qwe_\phi =
 \half \delta \qwe_{ \phi \ell}\up3 =
 \half \delta \qwe_{ \phi q }\up3 =
       \delta \qwe_{ \ell \ell }\up3 =
       \delta \qwe_{ \ell q}\up3 =
       \delta \qwe_{ q q }\up3 =
\inv{\lambda \epsilon} \delta \lambda = \qwe_{DW} g^2
\label{the.deltas}
\end{eqnarray}
which, in fact, leave $ \Delta( S , T , U ) $ invariant.

This result can also be obtained without using the equations of motion
on $ \ocal_{ DW } $. Adding a term $ \qwe_{DW} \ocal_{DW}/ \Lambda^2 $
to the effective Lagrangian
generates a quadratic term in the vector bosons, $ \qwe_{
DW} W^I_\mu \partial^2 W^{I\, \mu} $. When the quadratic part of the
vector-boson Lagrangian is re-diagonalized the $W$ and $Z$ masses
and the \vev\ $v$ are
modified, $ \delta M_W^2 / M_W^2 =  \delta M_Z^2 /  M_Z^2
=  \delta v /  v = g^2 \qwe_{DW} \epsilon $. The
Fermi constant is unaffected, $ \delta G_F = 0 $,
and the coupling of the $Z$ to the
left-handed fermionic current $J_L$ becomes $ - ( 1 + g^2
\qwe_{DW} \epsilon ) \sqrt{ g^2 + g'{}^2 } J_L \cdot Z $.
It is a tedious exercise (for which we used Ref.~\cite{bw}
after correcting a few typographical errors) to show that these
modifications correspond to (\ref{the.deltas}). This illustrates the fact that
(\ref{stu}) are consistent definitions of the heavy physics to the
oblique parameters.

In contrast, the naive definition of the oblique parameters in terms of the
vacuum polarization tensors, are not invariant under the replacement
(\ref{repl}).
Indeed, using an $ \su2 \times \ui $ basis,
\begin{eqnarray}
\svacpol  &=& - { 8 \pi \over M_Z^2 } \left[ \Pi_{3Y} \left( M_Z^2 \right) -
\Pi_{3Y} ( 0 ) \right] \nonumber\\
\tvacpol  &=& { 16 \pi \over \sin^2 ( 2 \theta_W ) M_Z^2 } \left[ \Pi_{11} (0)
-
\Pi_{33}(0) \right] \nonumber \\
\uvacpol  &=&
{ 16 \pi \over M_W^2 } \left[ \Pi_{11} \left( M_W^2 \right) - \Pi_{11} (0)
\right]
-
{ 16 \pi \over M_Z^2 } \left[ \Pi_{33} \left( M_Z^2 \right) - \Pi_{33} (0)
\right]
\label{naive}
\end{eqnarray}
from which, using (\ref{leff}), we obtain
\begin{equation}
\Dellin \tvacpol  = - \frac{4 \pi}{g^2 \; x} \alpha_\phi\up3 \epsilon + O (
\epsilon^2), \qquad
\Dellin \svacpol  = \frac{32 \pi y}{g^2 \; x} \qwe_{WB} \epsilon + O (
\epsilon^2 ), \qquad
\Dellin \uvacpol  = O ( \epsilon^2),
\label{stu.naive}
\end{equation}
but in this case (\ref{repl}) does {\em not} leave $\uvacpol$ invariant,
\begin{equation}
\Dellin \uvacpol  \rightarrow 16 \pi g'{}^2 \qwe_{DW} \; \epsilon + O (
\epsilon^2 )
\label{stu.naive.2}
\end{equation}
which illustrates the importance of using the definitions
(\ref{stu.physical}) for the oblique parameters.

It must be noted
that the operator $ \ocal_{DW} $ generates a $ p^4 $
contribution to the vacuum polarizations $ \Pi ( p ) $, and
within the linear approximation~\cite{peskin,roy} in $p^2$, this operator
will not affect the oblique parameters. This does not mean that
the effective lagrangian contributions to $\svacpol$, $\tvacpol$
and $\uvacpol$ within the linear approximation are unambiguous.
Consider, for example, the operator $ \left( \phi^\dagger D^\mu \phi
\right) \partial^\nu B_{\nu\mu} $ which contributes to
$ \Dellin \svacpol  $; using the equations of motion this
operator is equivalent to $ ( i g'/2 ) \left( 2 \ocal_\phi\up3 +
\ocal_\phi\up1 \right) $ --- plus a string of operators involving fermions
which do not contribute to (\ref{naive}) --- and $ \ocal_\phi\up3 $
contributes to $ \Dellin \tvacpol  $ only. In contrast, the
definitions (\ref{stu.physical}) present no such ambiguity
and should be used whenever an effective Lagrangian
computation is performed.

Using (\ref{stu}) or (\ref{stu.chiral}) and currently
available data we can derive limits on the scale of new physics $
\Lambda $.
The operators $ \ocal_{ \phi \ell} \up{1,3} $ and $ \ocal_{\phi e} $
modify the $Z$ coupling to the fermions and the corresponding
coefficients can be bounded using data from
LEP1~\cite{grzadkowski.wudka},
\begin{equation}
\epsilon \left|\qwe_{\phi \ell} \up{1,2} \right| < 0.0016, \qquad
\epsilon \left|\qwe_{\phi e} \right| < 0.0014 ;
\end{equation}
the operator $ \ocal_{\ell \ell}\up3 $ contributes to $e^+ e^-
\rightarrow \mu^+ \mu^- $ and it coefficient can be correspondingly
bounded~\footnote{There are many other operators that contribute to this
reaction, we assume there are no significant cancelations},
\begin{equation}
-0.105 < \epsilon \; \qwe_{\ell \ell } \up3 < 0.056 .
\end{equation}
Finally the limits on the oblique parameters are~\cite{pdg}
\begin{equation}
-0.0414 < { g^2 \, x \over 8 \pi } S < 0.0060, \qquad
-0.0875 < { g^2 \, x \over 4 \pi } T < 0.0951, \qquad
-0.0072 < { g^2 \over 16 \pi } U < 0.0020.
\end{equation}

Note however that the operators $ \ocal_{\ell \ell }\up3 ,\,
\ocal_{ \phi \ell} \up3 $ and $ \ocal_{ \phi W } $ also contribute to
$G_F$ and this can be used to impose better bounds on the corresponding
coefficients (again assuming no cancelations). Using $G_F$, $ \alpha $
and $ M_Z $ as input parameters the uncertainity in the predictions of
$ M_W $ requires $ \epsilon \left| \qwe_{\ell \ell }\up3 \right| ,\,
\epsilon \left| \qwe_{ \phi \ell} \up3 \right| ,\,
\epsilon \left| \qwe_{ \phi W }\right| \lesim 5 \times 10^{-4} $.
In the linear case we then have, to a good approximation,
\begin{eqnarray}
\Dellin T &\simeq& - \frac{4\pi}{g^2 \; x} \left( - 2\qwe_{\phi \ell}\up1+
\qwe_\phi\up3
+ 2\qwe_{\phi e} \right) \epsilon + O ( \epsilon^2 ), \nonumber\\
\Dellin S &\simeq& \frac{8\pi}{g^2 \; x} \left( -\qwe_{\phi e} +
2 x \qwe_{\phi \ell}\up1 + 4 y \qwe_{ W B } \right) \epsilon + O ( \epsilon^2),
\nonumber \\
\Dellin U &\simeq& \frac{16\pi}{g^2} \left( - 2 \qwe_{\phi \ell}\up1 +
\qwe_{\phi e}
\right) \epsilon + O ( \epsilon^2).
\end{eqnarray}
In the chiral case,
\begin{eqnarray}
\Delchir T &=& - \frac{4\pi}{g^2 \; x} \left( - 2\qwe_{\phi \ell}\up1+
\qwe_\phi\up3
+ 2\qwe_{\phi e} \right) \epsilon + O ( \epsilon^2 ), \nonumber\\
\Delchir S &=& \frac{8\pi}{g^2 \; x} \left( -\qwe_{\phi e} + 2 x \qwe_{\phi
\ell}\up1 + 4 y \qwe_{ W B } \right) \epsilon + O ( \epsilon^2), \nonumber \\
\Delchir U &=& \frac{16\pi}{g^2} \left( - 2 \qwe_{\phi \ell}\up1 + \qwe_{\phi
e} - 2 \qwe_{ W W } \right) \epsilon + O ( \epsilon^2).
\end{eqnarray}

Using these expressions and the above experimental constraints
we find the following bounds,
\begin{equation}
\epsilon \left| \qwe_\phi\up3 \right| \lesim 0.1, \qquad
\epsilon \left| \qwe_{ WB} \right| \lesim 0.02 , \qquad
\epsilon \left| \qwe_{W W } \right| \lesim0.006 ,
\end{equation}
(where the last is relevant only for the chiral case).

In the linear case the natural size~\cite{aew} for the
coefficients are $ \left| \qwe_\phi\up3 \right| \lesim 1 $ and
$ \left| \qwe_{ WB} \right| \lesim g g'/ (
4 \pi )^2 $.
The limit on $ \qwe_\phi\up3 $ implies $ \Lambda \gesim 550 \gev $
while, from $ \qwe_{WB} $, $ \Lambda \gesim 50 \gev $.
This disparity is due to the fact that $ \ocal_\phi\up3 $ can be
generated at tree level by the heavy physics,
while $ \ocal_{WW} $ is necessarily loop
generated~\cite{aew}. The $ 550 \gev $ limit refers to
the mass of a heavy scalar or vector boson whose interactions violate
the custodial symmetry~\cite{custodial}.

For the chiral case the natural sizes~\cite{nda}
are $ \left| \qwe_\phi\up3 \right| \lesim1 , \
\left| \qwe_{ WB} \right| \lesim g \, g'$, and $
 \left| \qwe_{W W }\right| \lesim g^2 $; moreover we also have $
\Lambda \sim 4 \pi v \sim 3 \tev $ so that
$ \epsilon \lesim 1/ ( 4 \pi )^2 $. The
above limits are not sufficiently precise to provide useful information
in this case; for example, the limit on $ \qwe_{WW} $ implies $ \Lambda
\gesim 1.5 \tev $.

Finally we note a peculiarity of the
parameter $U$: the heavy physics contributions generated by $
\ocal_{\phi \ell}\up1 $ and $ \ocal_{\phi e} $ were measured at LEP1 and
are known to be small; this means that in the linear
case current data implies $ \Delta U \sim 0 $ ($ U \sim \epsilon^2 \lesim 0.01
$
for $ \Lambda > 550 \gev $).
In contrast there are no severe bounds on the contributions
generated by $ \ocal_{WW} $; in the chiral case we therefore have $ |
\Delchir U | \sim 32 \pi | \qwe_{WW} | \epsilon \lesim 2/\pi $.
Should a future
measurement produce a deviation of order $0.1 $ in the measurement of
$U$, this observation would not
only indicate the presence of new physics, but would strongly disfavor
the existence of light Higgs-like scalars. Note, however, that a bound $
\Delta U \lesim 0.1 $ does {\em not} imply the presence of light
scalars since this could also occur within the chiral case for a
sufficiently large $ \Lambda $.

\acknowledgements
This work was partially supported by CONACyT (M\'exico) and by the U.S.
Department of Energy under contract DE-FG03-94ER40837.


\begin{references}

\bibitem{peskin}
M. E. Peskin and T. Takeuchi, \prd{46}{381}{1992}.\\
G.~Altarelli and R.~Barbieri, Phys. Lett. {\bf B253} (1991) 161.\\
G.~Altarelli, R.~Barbieri, and S.~Jadach, Nucl. Phys. {\bf B369} (1992) 3.\\
D.~C.~Kennedy and P.~Langaker, Phys. Rev. Lett. {\bf 65} (1990) 2967; {\bf 66}
(1991) 395(E); Phys. Rev. {\bf D44} (1991) 1591.\\
W.~J.~Marciano and J.~L.~Rosner, Phys. Rev. Lett. {\bf 65} (1990) 2963; {\bf
68} (1992) 68(E).
%
\bibitem{new.stu} See for example,\\
P. Langacker, M.-X. Luo and A.K. Mann, \rmp{64}{87}{1992}.\\
K. Lane, in: proc of the  {\sl 27th Int. Conf. on High Energy Physics (ICHEP)},
Glasgow, Scotland, Jul 20-27, 1994. Edited by P.J. Bussey and I.G. Knowles (IOP
press, 1995).\\
K. Hagiwara {\it et al.}, Z. Phys. {\bf C64} (1994) 559; {\bf C68} (1995)
352(E).\\
K. Hagiwara, D.~Haidt, and S.~Matsumoto, hep-ph/9706331.\\
S. Alam, S. Dawson, R. Szalapski, \prd{57}{1577}{1998}.
%
\bibitem{roy}
G. Bhattacharyya, S. Banerjee, and P. Roy, \prd{45}{R729}{1992}; {\bf D46}
(1992) 3215(E).\\
A. Kundu and P. Roy, \ijmpa{12}{1511}{1997}.
%
\bibitem{arzt}
C. Arzt, \plb{342}{189}{1995}.\\
H.~Georgi, \npb{361}{339}{1991}; \npb{363}{301}{1991}.\\
U.~G.~Meissner, Rep. Prog. Phys. {\bf 56} (1993) 903.
%
\bibitem{hagiwara}
K. Hagiwara {\it et al.}, \prd{48}{2182}{1993}.\\
K. Hagiwara, S.~Matsumoto, and R.~Szalapski, \plb{357}{411}{1995}.
%
\bibitem{wudka}
J. Wudka, \ijmpa{9}{2301}{1994}.
%
\bibitem{sm.stu}
F. Halzen, P.~Roy, and M.~L.~Strong, Phys. Lett. {\bf B277} (1992) 503.\\
F. Halzen, B.~A.~Kniehl, and M.~L.~Strong, Z. Phys. {\bf C58} (1993) 119.\\
ZFITTER, D.~Bardin {\it et al.}, Z. Phys. {\bf C44} (1989) 493.\\
D.~Bardin {\it et al.}, \npb{351}{1}{1991}.\\
D.~Bardin {\it et al.}, \plb{255}{290}{1991}.\\
See also Refs.~\cite{new.stu}.
%
\bibitem{leff.refs}
H. Georgi and S. Weinberg, \physa{96}{327}{1979}.\\
A. Pich, in  proceedings of the {\it 5TH Mexican School of Particles and
Fields}, edited by J.L. Lucio and M. Vargas (Amer. Inst. Phys., New York,
1994).\\
G. Ecker, in proceedings of the {\it International Workshop on Hadron Physics
96}, edited by E. Ferreira {\it et al.} (World Scientific, 1997).
%
\bibitem{leff.linear}
J. M. Cornwall, D.~N.~Levin, and G.~Tiktopoulos, \prd{10}{1145}{1974};
\prd{11}{972(E)}{1975}.\\
B.~W.~Lee, C.~Quigg, and H.~Thacker, \prd{16}{1519}{1977}.\\
M.~Veltman, Acta Phys. Pol. {\bf B8} (1977) 475.\\
C.J.C. Burges and H.J. Schnitzer, \npb{228}{464}{1983}.\\
C.N. Leung, S.T. Love and S. Rao, \zphys{31}{433}{1986}.\\
W. Buchm\"uller and D. Wyler, Ref. \cite{bw}.
%
\bibitem{leff.chiral}
S. Weinberg, \physa{96}{327}{1979}.\\
M.~Chanowitz and M.~K.~Gaillard, \npb{261}{379}{1985}.\\
M.~Chanowitz, M.~Golden and H.~Georgi, \prd{36}{1490}{1987}.\\
A.~Dobado and M.~J.~Herrero, \plb{228}{495}{1989}.\\
H.~Georgi, \npb{361}{339}{1991}; \npb{363}{301}{1991}.\\
B.~Holdom, \plb{259}{329}{1991}.
%
\bibitem{bw}
W. Buchm\"uller and D. Wyler, \npb{268}{621}{1986}.
%
\bibitem{appelquist.wu}
T. Appelquist and G.-H. Wu, \prd{48}{3235}{1993}; \prd{51}{240}{1995}.
%
\bibitem{grzadkowski.wudka}
B. Grzadkowski and J. Wudka, \plb{364}{49}{1995}.
%
\bibitem{pdg}
R. M. Barnett {\it et al.}, \prd{54}{1}{1996}.
%
\bibitem{aew}
C. Arzt, M.~B.~Einhorn and J.~Wudka, \npb{433}{41}{1995}.
%
\bibitem{custodial}
J. Ellis {\it et al.}, \npb{182}{529}{1981}.
%
\bibitem{nda}
S. Weinberg, \physa{96}{327}{1979}.\\
H. Georgi and A. Manohar, \npb{234}{189}{1984}.\\
H. Georgi, \plb{298}{187}{1993}.
%
\end{references}
\end{document}